\documentclass[letterpaper]{jpconf}

\usepackage{graphicx}

\begin{document}

\title{Precision Tests of QCD Using Final State Jets and Particles}

%      \begin{footnote}{Talk given at the Lake Louise Winter Institute
%       February 15-20, Alberta, Canada}
%     \end{footnote}

\author{Mihajlo Mudrini\'c}

\address{
DESY, 22607 Hamburg, Germany\\
and "Vinca" Institute of Nuclear Science, Serbia\\
\vspace*{2mm}
\rm on behalf of the H1 and ZEUS Collaborations\\
}

\ead{mihajlo.mudrinic@desy.de}

\begin{abstract}

The data from the HERA experiments H1 and ZEUS allows a precise extraction of the strong coupling constant $\alpha_S$ 
with the highest experimental precision (sub 1\%). A review of recent measurements of jet cross sections in 
neutral current deep inelastic scattering (NC DIS) at HERA is presented and compared with 
theoretical NLO QCD predictions. The latest determinations of $\alpha_S$ in a large range of $Q^2$ are shown.

\end{abstract}

\section{Introduction}

In $e^{\pm} p$ collisions at HERA, according to the virtuality 
$Q^2$ of the exchanged boson we distinguish two processes, DIS and photoproduction.
In DIS process ($Q^2>1$~GeV$^2$) a highly virtual boson interacts with a parton carrying a 
momentum fraction of the proton. Among the spray of particles emerging from 
 a high-energy reaction we can recognize collimated subsystems of hadrons, so called jets.
Jets in DIS at HERA result from the scattered quark and from additional QCD radiation either in the initial 
or the final state. For quantitaive measurements we need to define jets using a jet algoritam 
which prescribes how to combine object (partons, hadrons, energy deposit, ..) close in phase space to jets. 
In the analyses presented here jets are defined using the $k_T$ clustering algorithm applied in the Breit frame. 
The associated cross sections are collinear and infrared safe and therefore well suited for comparison with predictions 
from fixed order QCD calculations. Jet production in $e^{\pm} p$ collisions proceeds 
via the Born, boson-gluon fusion and QCD Compton processes. In the Breit frame, where the virtual boson and the
proton collide head on, siginificant transverse momenta ($E_T$) are produced at leading order 
(LO) in $\alpha_S$ by the boson-gluon fusion and QCD Compton processes. In the pQCD, the jet cross section depends 
on the strong coupling constant $\alpha_S$ as the expansion parameter for the perturbation series and on the parton
densities in the proton (PDFs). In regions where the PDFs are well constrained, the
jet data allow a test of the  general  aspects of pQCD. In regions where the PDFs are not so well constrained, 
jet cross sections can be incorporated into global QCD fits to help further constrain them. 
This contribution presents the latest jet production
studies made by the H1 and ZEUS Collaborations and 
the resulting determinations of $\alpha_S$. Such determinations
allow a stringent test of the running of $\alpha_S$ predicted by pQCD.

%-------------------------------------------------------------------------
\section{Jet Measurements in DIS}

\begin{figure}[h]
\begin{center}
\includegraphics[width=10cm,height=9cm]{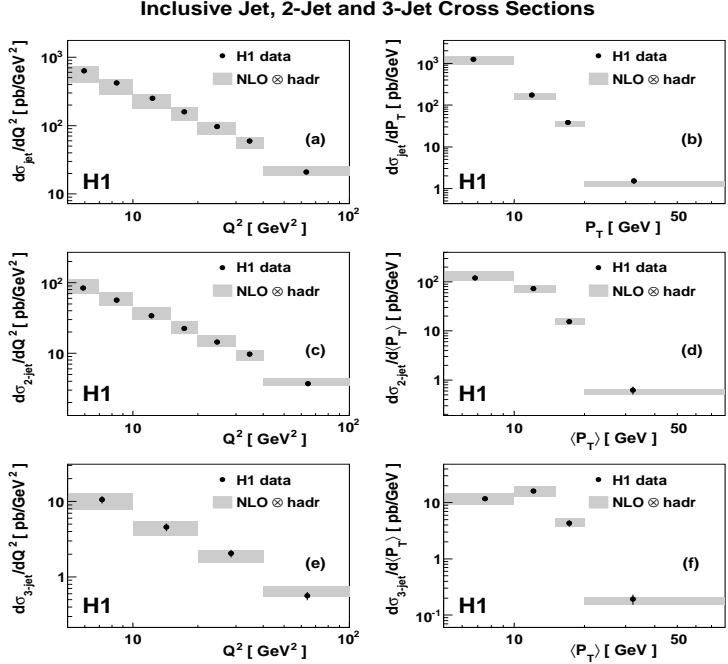}
\caption{
Inclusive jet cross sections $d\sigma_{jet} / dQ^2$ (a) and $d\sigma_{jet} / dP_T$ (b), 2-jet cross sections
 $d\sigma_{2-jet} / dQ^2$ (c) and $d\sigma_{2-jet} / d\langle P_T \rangle$ (d) and 3-jet cross sections $d\sigma_{3-jet} / dQ^2$  (e) and  $d\sigma_{3-jet} / d \langle P_T \rangle$ (f), compared with NLO QCD predictions corrected for hadronisation. The error bars show the total experimental uncertainty, formed as the quadratic sum of the statistical
and systematic uncertainties. The points are shown at the average values of $Q^2$, $P_T$ or $\langle P_T \rangle$ within each bin. The NLO QCD predictions are shown together with the theoretical uncertainties
associated with the scale uncertainties and the hadronisation (grey band).
}
\label{fig:lowQ2} 
\end{center}
\end{figure}

Jet measurements in DIS were recently performed by the H1 Collaboration in two 
kinematic regimes. The low $Q^2$ data~\cite{bib:LowQ2}
corresponding to \mbox{$5 < Q^2 < 100~\rm GeV^2$} use a HERA-I sample of $44$~pb$^{-1}$,  
whereas the high $Q^2$ data~\cite{bib:HighQ2},
corresponding to \mbox{$150 < Q^2 < 15000~\rm GeV^2$}, are based on nearly the full H1 
data sample of about   $400$~pb$^{-1}$. The inclusive jet cross sections are measured in the low $Q^2$ 
regime by requiring \mbox{$E_T>5$~GeV} and $-1.0 < \eta^{\rm Lab} < 2.5$. Comparison of inclusive jet cross sections, 
2-jet cross sections and 3-jet cross sections for the low $Q^2$ regime with NLO QCD predictions 
are shown on Figure \ref{fig:lowQ2}.

\begin{figure}[h]
\begin{minipage}{0.32\linewidth} % A minipage that covers half the page
\centering
\includegraphics[width=4.9cm,height=5.5cm]{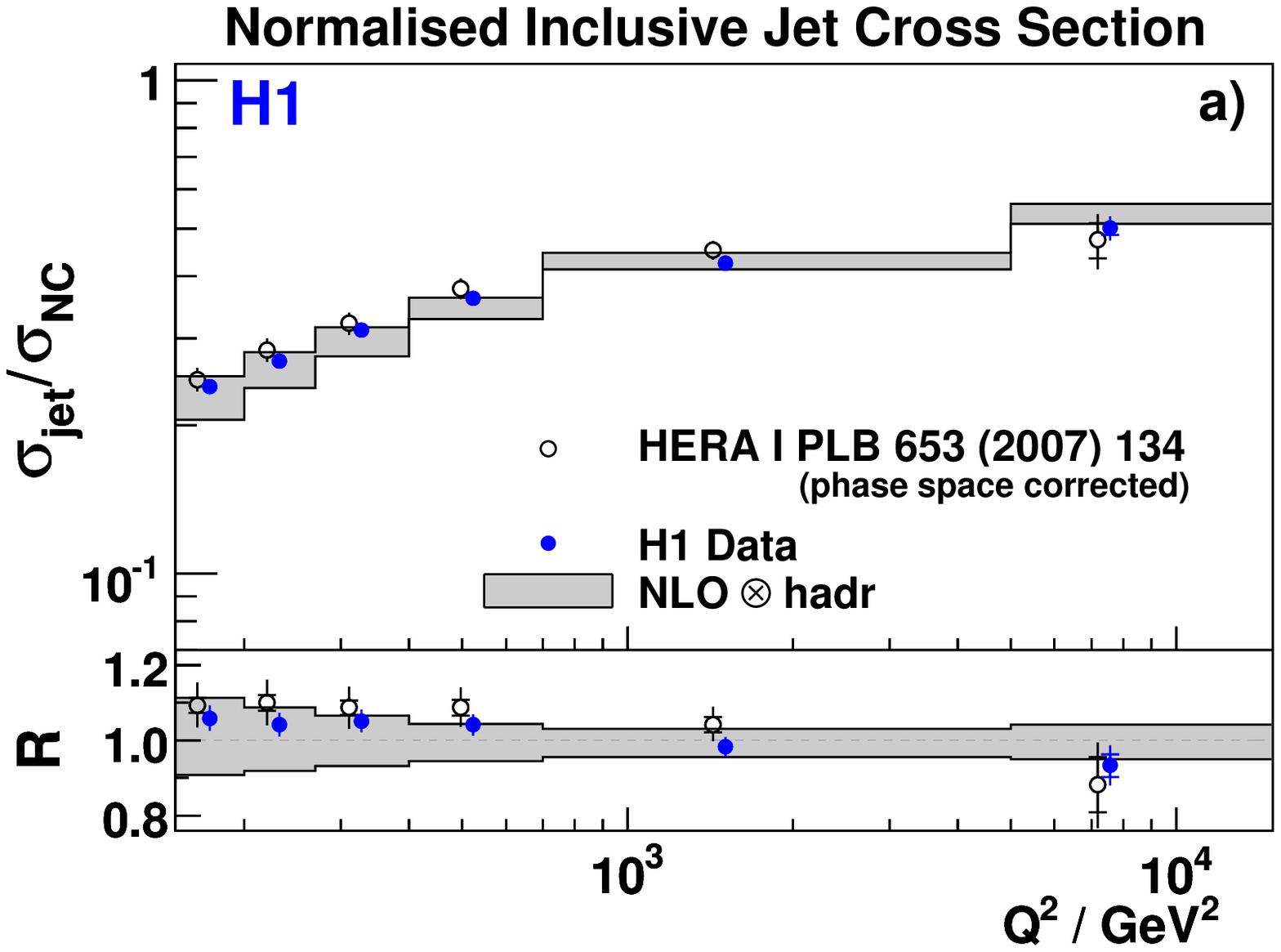}
\end{minipage}
%\hspace{1pc} % To get a little bit of space between the figures
\begin{minipage}{0.32\linewidth}
\centering
\includegraphics[width=4.9cm,height=5.5cm]{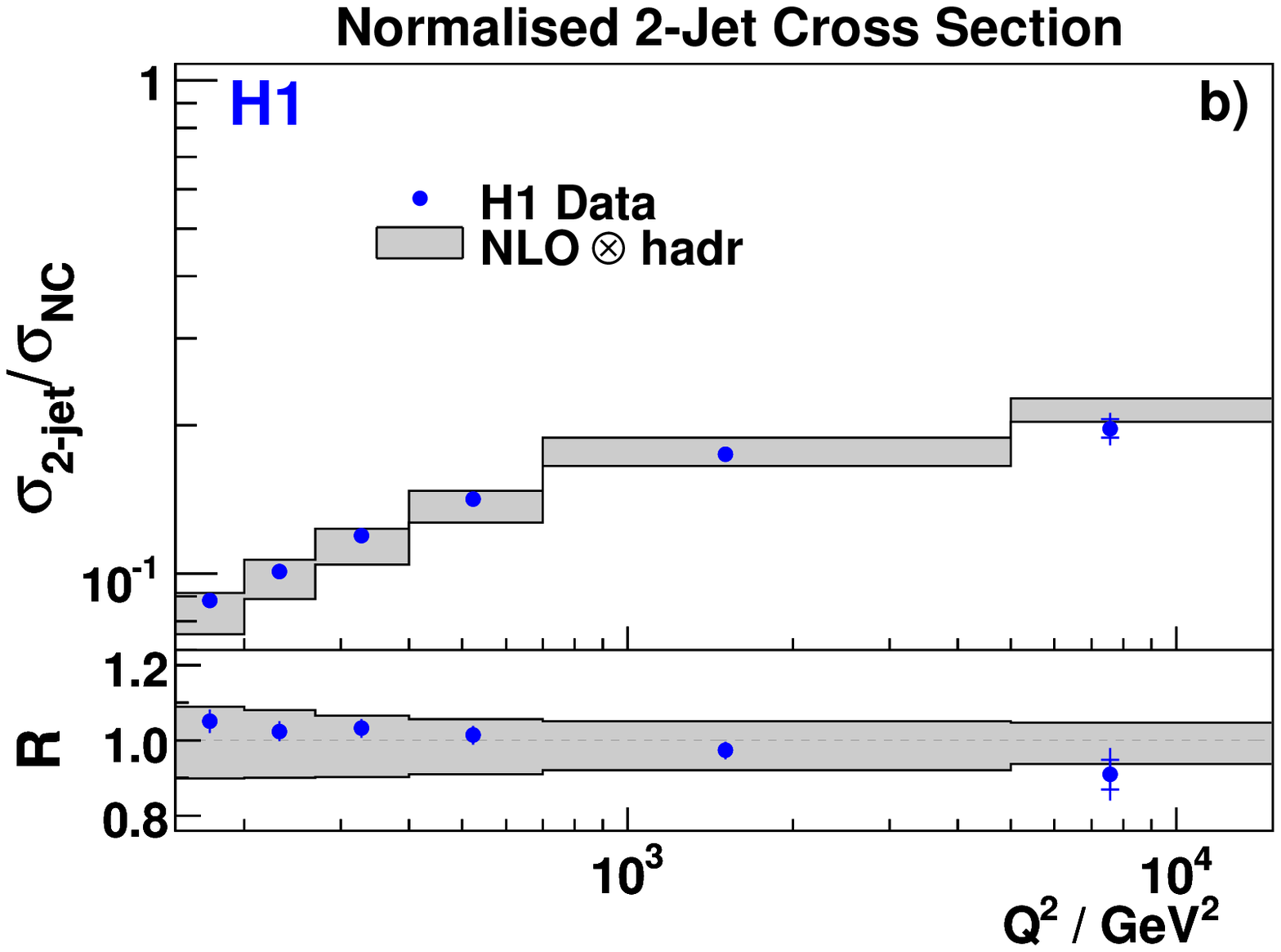}
\end{minipage}
%\hspace{1pc} % To get a little bit of space between the figures
\begin{minipage}{0.32\linewidth}
\centering
\includegraphics[width=4.9cm,height=5.5cm]{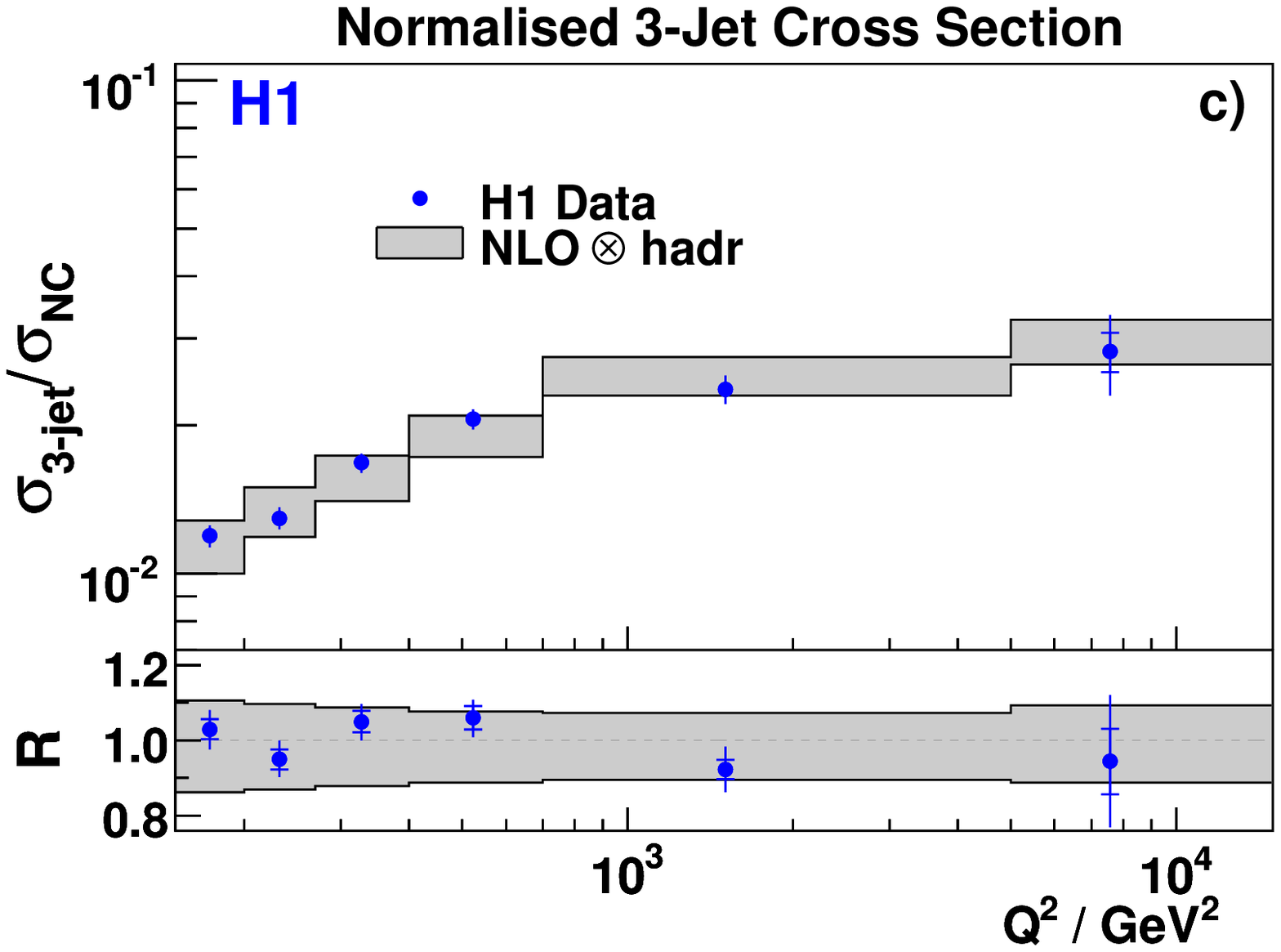}
\end{minipage}
\caption{Normalised inclusive (a), 2-jet (b) and 3-jet (c) cross sections
 in NC DIS for high $Q^2$ regime, measured as a function of $Q^2$. The points are shown at the average
value of $Q^2$ within each bin.}
\label{fig:highQ2}
\end{figure}

\begin{figure}[h]
\begin{minipage}{20pc}
\includegraphics[width=20pc,height=6.5cm]{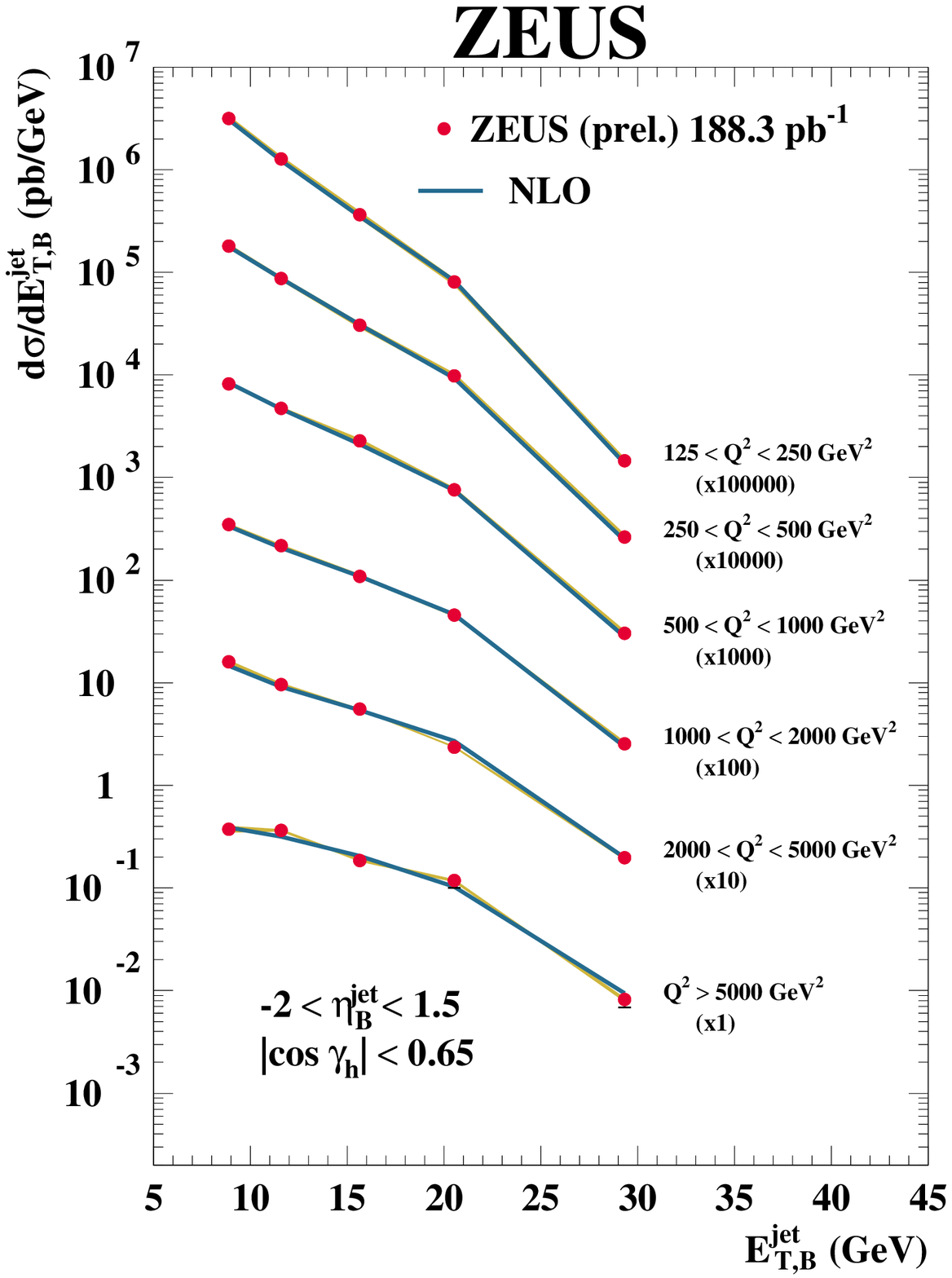}
\end{minipage}
\begin{minipage}{20pc}
\includegraphics[width=20pc,height=6.5cm]{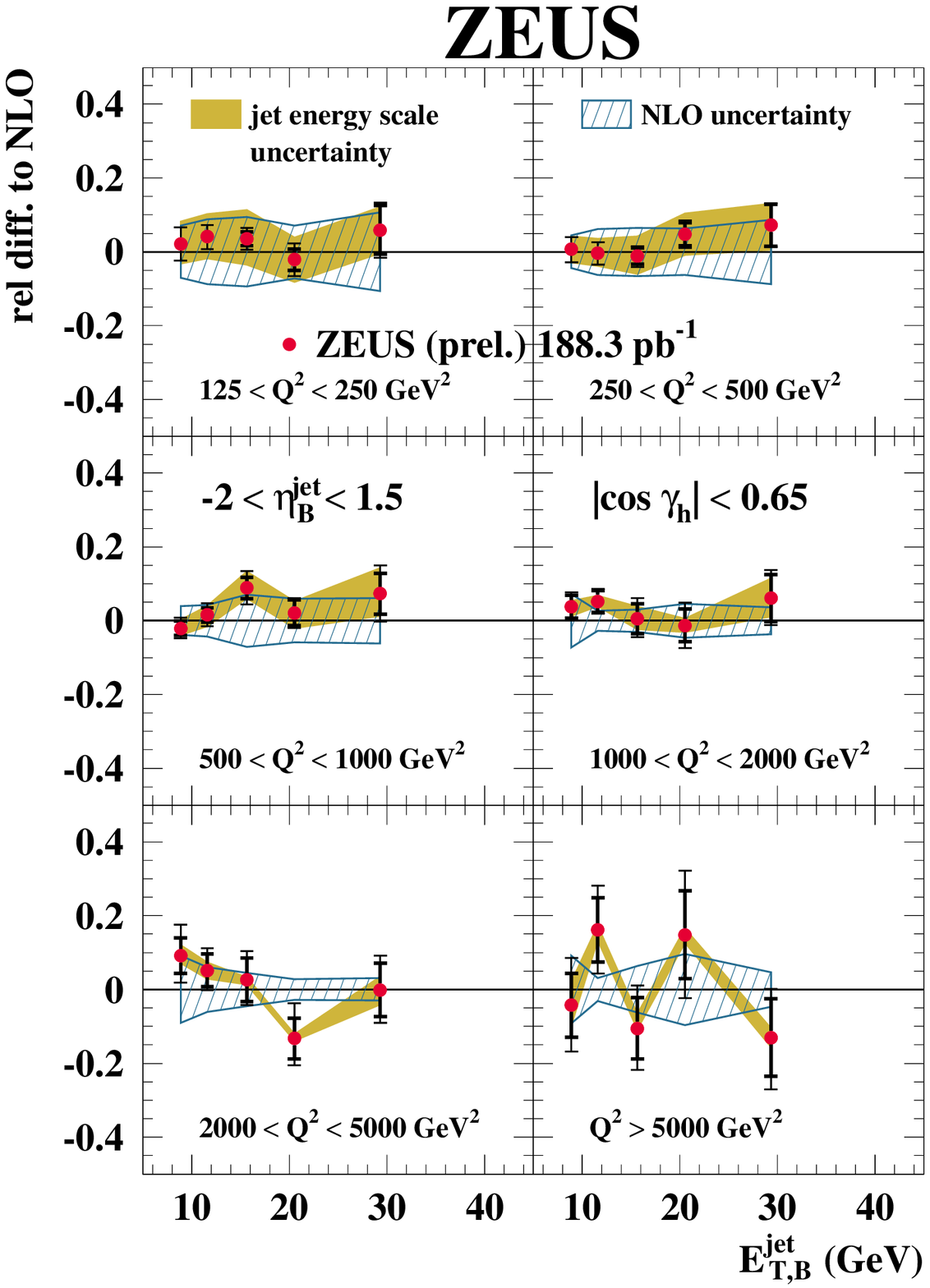}
\end{minipage}
\caption{The measured differential cross-section $d\sigma / d E_{T,B}^{jet}$ for inclusive-jet production 
in different regions of $Q^2$ (left). The relative difference between the measured differential cross-sections
 $d\sigma / d E_{T,B}^{jet}$ and NLO QCD calculation (right).
}
\label{fig:ZeushighQ2} 
\end{figure}

At high $Q^2$ the cross sections are measured for inclusive jets with $7 < E_{T} < 50$~GeV and for
2-jet and 3-jet events containing jets with $5 < E_{T} < 50$~GeV
and $-0.8 < \eta^{\rm Lab} < 2.0$. The jet cross sections at high $Q^2$ are normalised to the inclusive 
DIS cross sections in order to reduce the sensitivity to the absolute 
normalisation uncertainties. The normalised jet cross sections as a function of 
$Q^2$  are shown in Figure \ref{fig:highQ2}. One of the main sources of experimental 
uncertainties remains to be the uncertainty on the absolute calibration of the hadronic energy scale, 
which results in an uncertainty on the cross sections of about 1 to 5\%. The detector correction 
factors show an uncertainty due to the MC model dependence which amounts typically 
to $1$ to 10$\%$. Differential inclusive-jet cross sections have been measured in NC DIS 
$ep$ for boson virtuality  \mbox{$ Q^2 > 125~\rm GeV^2$} with ZEUS detector at HERA~\cite{bib:ZEUSAlpha}. 
The measurement of the differential cross section and relative difference between the measurement and NLO QCD
calculation are shown on Figure \ref{fig:ZeushighQ2}.

%%%%%%%%%%%%%%%%%%%%% FIG.1 %%%%%%%%%%%%%%%%%%%%%%%%%%%%%%%%%%%%%%%%%%

%%%%%%%%%%%%%%%%%%%%% FIG.2 %%%%%%%%%%%%%%%%%%%%%%%%%%%%%%%%%%%%%%%%%%

%%%%%%%%%%%%%%%%%%%%% FIG.3 %%%%%%%%%%%%%%%%%%%%%%%%%%%%%%%%%%%%%%%%%%

The H1 jet cross sections at low and high $Q^2$ are used to extract $\alpha_S$ ~\cite{bib:LowQ2}.
The experimental uncertainty of $\alpha_S$ is 
defined by that change in $\alpha_S$ which increases the minimal $\chi^2$ by one 
unit. 
The $\alpha_S$ is extracted individually from the inclusive jets at low 
$Q^2$ and from the inclusive, 2-jet and 3-jet measurement at high $Q^2$. The experimentally 
most precise determination of $\alpha_S(M_Z)$ is derived from the combined fit to 
all three observables at high $Q^2$:

$$
\alpha_S(M_Z)=0.1168\pm 0.0007~( {\rm exp.})^{+0.0046}_{-0.0030}~({\rm th.})\pm{0.0016}~({\rm PDF}).
$$

\begin{figure}[h]
\begin{minipage}{18pc}
\includegraphics[width=18pc,height=6.9cm]{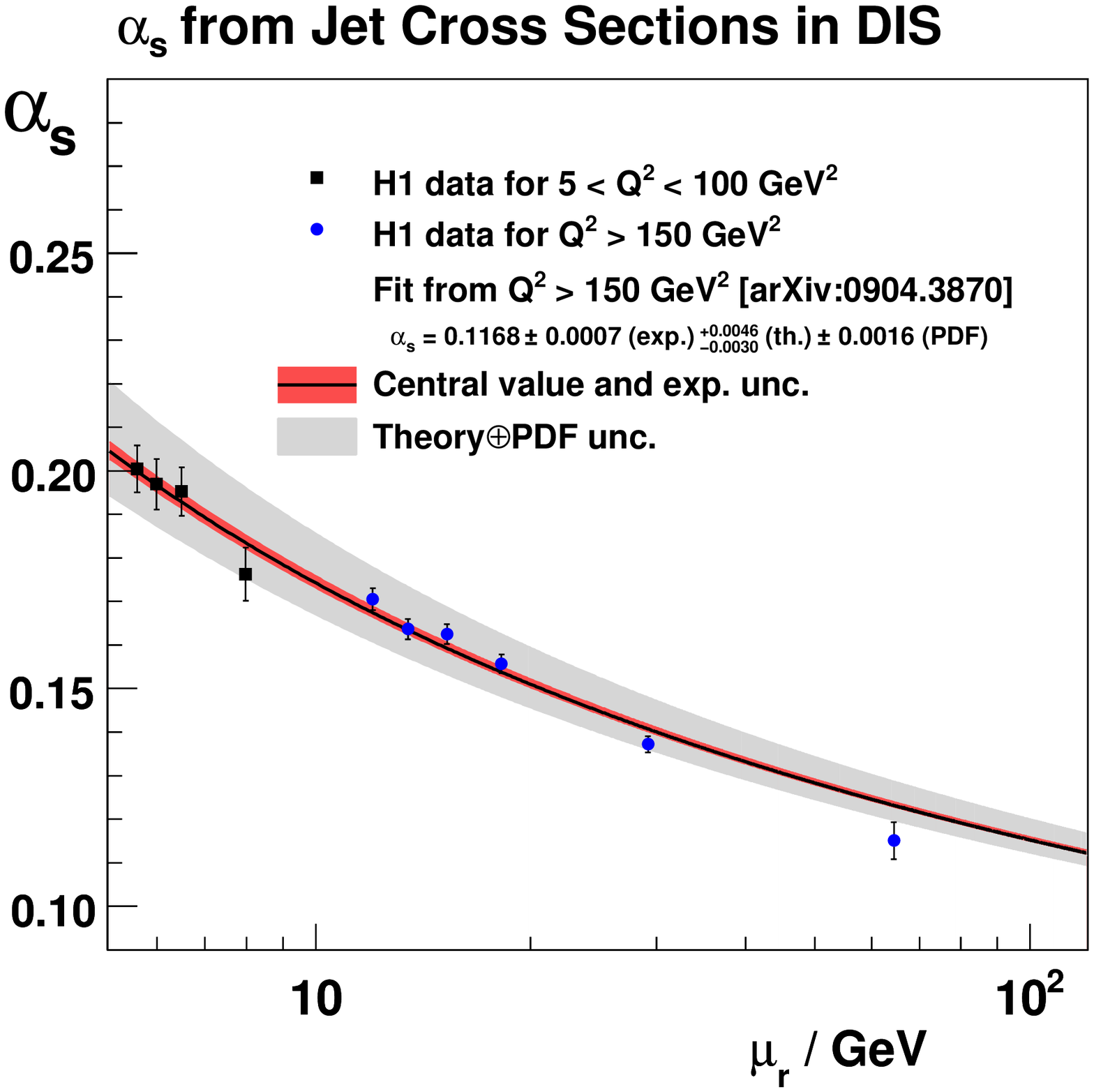}
\end{minipage}
\hspace{1pc}
\begin{minipage}{18pc}
\includegraphics[width=18pc,height=6.5cm]{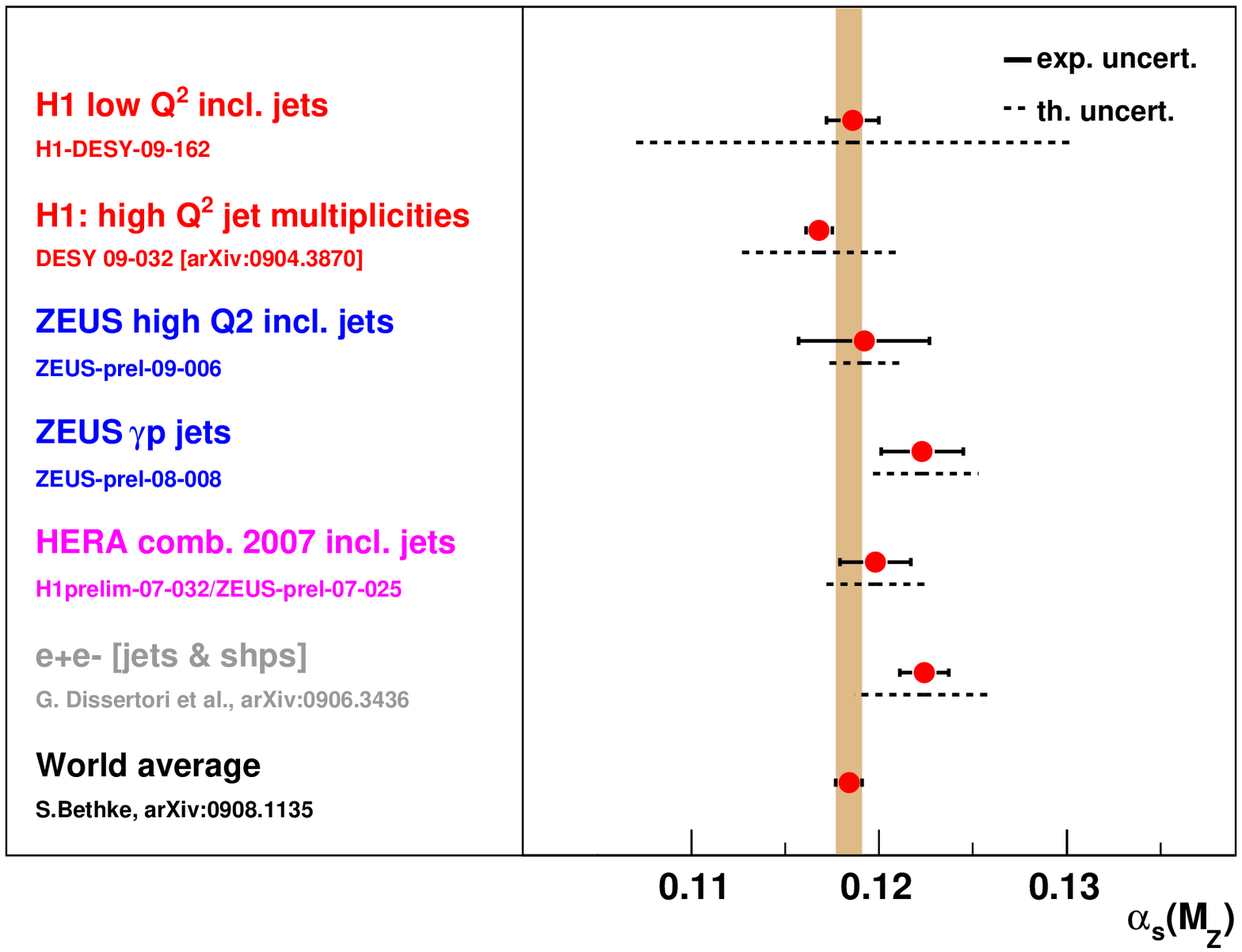}
\end{minipage}

\caption{The running of $\alpha_S(Q)$ (left) and different
         recent determinations of $\alpha_S(M_Z)$ from HERA, compared with
         a LEP measurement and the world average (right).}
\label{fig:alphas} 
\end{figure}

 The theory uncertainty is estimated by the \textit{offset method}, adding in 
quadrature the deviations due to various choices of scales and hadronisation 
corrections. The largest contribution is the theoretical uncertainty arising from 
terms beyond NLO which amounts to 3$\%$. The PDF uncertainty, estimated using 
CTEQ6.5, amounts to $1.5\%$. 
 The value extracted at low $Q^2$, 
$
\alpha_S(M_Z) = 0.1160 ~\pm 0.0014~({\rm exp.}) ^{+0.0093}_{-0.0077}~({\rm th.}) \pm 0.0016~({\rm PDF})
$
is compatible with high the $Q^2$ value, but the
uncertainty arising from the renormalisation scale variation reach 10$\%$.  
The measurement of the strong coupling in a  
large $Q^2$ range allows to test the value  
$\alpha_S(\mu_r)$ where $\mu_r = \sqrt{(Q^2 + P_{T,pbs}^2)/ 2}$ is renormalisation scale 
running between 6 and 70 GeV as shown in 
Figure \ref{fig:alphas}.

The ZEUS Collaboration has determined $\alpha_S$ from the inclusive jet
cross section in DIS. 
The analysis is based on the HERA-II data sample corresponding to
 an integrated luminosity of 188.3~pb$^{-1}$.
In order to reduce the theory uncertainty on
$\alpha_S$ ~\cite{bib:Jones}, only the data at $Q^2 > 500 GeV^2$ are used in the fit,
resulting to value of: 
$$
\alpha_S(M_Z)=0.1192\pm 0.0009~({\rm stat.})^{+0.0035}_{-0.0032}\ ({\rm exp.})
^{+0.0020}_{-0.0021}\ ({\rm th.}).
$$

\section{Conclusions}
Numerous measurements of jet production in DIS
have been made over a wide kinematic range at
HERA. In this overview only a few recent results
could be given. In general, the data are well described
by NLO QCD predictions, and the small
experimental uncertainty in the high $Q^2$ regime
allows the exctraction of values of $\alpha_S$ with high experimental
precision, which is not yet matched by the theory error. 
The results for jets at HERA are precise and 
competitive with those from $e^+ e^-$ 
data~\cite{bib:BETHKE} and are in good
agreement with world averages~\cite{bib:BETHKE,bib:WORLD}.

%%%%%%%%%%%%%%%%%%%%%%%%%%%%%%%%%%%%%%%%%%%%%%%%%%%%%%%%%%%%%%%

\end{document}